\documentclass[11pt]{article}
\usepackage{graphicx}
\usepackage{amssymb}
\usepackage{graphics}

\input{tcilatex}
\begin{document}

\title{Ultracold neutron accumulation in a superfluid-helium converter with
magnetic multipole reflector}
\author{O. Zimmer$^{1}$ and R. Golub$^{2}$ \\
%EndAName
$^{1}$Institut Laue Langevin, 38042 Grenoble, France\\
$^{2}$Department of Physics, North Carolina State University, Raleigh, NC
27695, USA}
\maketitle

\begin{abstract}
\noindent We analyze accumulation of ultracold neutrons (UCN) in a
superfluid-helium converter vessel surrounded by a magnetic multipole
reflector. We solved the spin-dependent rate equation, employing formulas
valid for adiabatic spin transport of trapped UCN in mechanical equilibrium.
Results for saturation UCN densities are obtained in dependence of order and
strength of the multipolar field. The addition of magnetic storage to
neutron optical potentials can increase the density and energy of the low
field seeking UCN\ produced and serves to mitigate the effects of wall
losses on the source performance. It also can provide a highly polarized
sample of UCN without need to polarize the neutron beam incident on the
converter. This work was performed in preparation of the UCN source project
SuperSUN at the ILL.\bigskip

\bigskip\ source, neutron EDM,
neutron decay
\end{abstract}

\section{Introduction}

Mirror reflection of neutrons is an effect of the neutron optical potential
which is mainly due to coherent s-wave scattering of neutrons by atomic
nuclei in condensed matter \cite{Fermi/1936}. Ultracold neutrons (UCN),
which were first produced in Dubna \cite{Luschikov/1969} and in Munich \cite%
{Steyerl/1969}, have energy sufficiently low to become totally reflected
under any angle of incidence. This peculiar property enables
experimentalists to store them in "neutron bottles" made of suitable
materials with small cross sections for neutron absorption and providing
well depths up to about $300$ neV \cite{Golub/1991,Ignatovich/1990}. Storage
time constants of many hundreds of seconds and the possibility to employ
also magnetic fields and gravity for trapping and manipulation have made UCN
a versatile tool to investigate various phenomena of fundamental physics
complementary to experiments at high-energy particle accelerators \cite%
{Dubbers/2011,Musolf/2008,Abele/2008}.

Among recent experiments with UCN feature a first demonstration of gravity
resonance spectroscopy with the goal to search for deviations from Newton's
gravity law at the micrometer length scale \cite{Jenke/2011}, searches for
\textquotedblleft mirror dark matter\textquotedblright\ \cite%
{Serebrov/2008,Ban/2007}, a test of Lorentz invariance \cite{Altarev/2009},
searches for axion-like particles \cite%
{Jenke/2014,Serebrov/2010,Zimmer/2010b}, a demonstration of the effect of
accelerated matter on the neutron wave \cite{Frank/2008} and of the
stability of the Berry geometrical phase for spin 
%TCIMACRO{\U{bd} }%
%BeginExpansion
$\frac12$
%EndExpansion
particles under the influence of noise fluctuations \cite{Filipp/2009}.
Earlier work with UCN on the geometrical phase was published in \cite%
{Richardson/1989,Richardson/1988}, while its first demonstration with cold
neutrons can be found in \cite{Bitter/1987}.

Long standing are efforts to improve the accuracy of the weak axial-vector
and vector coupling constants of the nucleon derived from precise values of
the neutron lifetime \cite%
{Arzumanov/2012,Pichlmaier/2010,Paul/2009,Serebrov/2008b,Nico/2005} and the
beta-asymmetry \cite{Mendenhall/2013,Plaster/2012,Mund/2012,Abele/2002}.
Among other applications these values are crucial input for calculations of
weak reaction rates in big-bang nucleo-synthesis and stellar fusion \cite%
{Coc/2007,Lopez/1999}, and of the efficiency of neutrino detectors \cite%
{Mention/2011}. Also long standing is the search for a non-vanishing neutron
electric dipole moment (EDM), which would violate the symmetries of parity
(P) and time reversal (T) and thus via the CPT theorem also the combined
symmetries of charge conjugation and parity (CP). This search was proposed
already in 1950 by Purcell and Ramsey \cite{Purcell/1950} and has become a
prominent route to investigate new mechanisms of CP violation beyond the
standard model's complex phase of the weak quark mixing CKM matrix, and the
matter-antimatter asymmetry in the universe \cite{Pospelov/2005}. At the
present best level of sensitivity, still no EDM was observed \cite%
{Baker/2006}. Several projects are in preparation or underway with the goal
to gain at least an order of magnitude in sensitivity \cite%
{Serebrov/2014,Altarev/2012,Masuda/2012a,Grinten/2009,Altarev/2009b,Serebrov/2009b,Lamoreaux/2009,EDM@SNS/2004,Golub/1994}%
.

Most studies with UCN are counting statistics limited and will strongly
profit from new UCN sources which are currently being developed in various
laboratories around the world \cite%
{Karch/2014,Lauss/2014,Piegsa/2014,Lauer/2013,Saunders/2013,Masuda/2012,Serebrov/2009,Korobkina/2007,Frei/2007,Trinks/2000}%
. They are all based on the "superthermal" UCN production scheme proposed in
1975 by one of the authors together with Mike Pendlebury \cite{Golub/1975},
using either superfluid $^{4}$He or solid deuterium as a medium for neutron
conversion. Early milestones in the development of the latter were published
in \cite{Serebrov/1994,Golub/1983-1,Altarev/1980}. Here we focus on UCN
production in a converter of superfluid $^{4}$He installed at the end of a
neutron guide, wherein neutrons with energy $1$ meV, respectively,
wavelength $0.89$ nm may loose nearly their entire energy in single
scattering events. At low temperature only few excitations are present in
the helium that are able to scatter UCN back to higher energies. With the
vanishing absorption cross section of $^{4}$He it becomes possible to
accumulate UCN within a converter with reflective boundaries before
releasing them to an experiment at room temperature. While an earlier
attempt to bring this technique to life was hampered by extraction losses
(nonetheless producing record UCN densities for its time) \cite%
{Kilvington/1987}, a more efficient method was developed recently by one of
the authors together with his co-workers \cite%
{Piegsa/2014,Zimmer/2011,Zimmer/2010,Zimmer/2007}. UCN are extracted through
a cold UCN valve and a short vertical UCN guide section, superseding lossy
separation window, screens and gaps for thermal insulation between the
converter and the UCN guide of the earlier scheme. Work is in progress to
bring the technique to maturity for a new user facility at the ILL, and in
particular for performing a neutron lifetime experiment using magnetic
trapping \cite{Leung/2014,Leung/2009,Zimmer/2000}. Also other groups have
recognized the potential advantages of a superfluid helium converter feeding
UCN to experiments at room temperature \cite{Masuda/2012,Serebrov/2009}, and
in some experiments this type of converter is employed in situ \cite%
{Grinten/2009,EDM@SNS/2004,Huffman/2000}.

The efficiency of a UCN accumulator at an external neutron beam relies on
loss rates being sufficiently low. Most critical are those losses which
occur when UCN hit the walls of the converter vessel. They are proportional
to the frequency of collisions and thus depend on the size and shape of the
converter vessel. From transmission measurements through superfluid $^{4}$He
at $1.25$ K a mean free path of $17$ m was derived for the $0.89$ nm
neutrons most effective for UCN production \cite{Sommers/1955}. Hence, the
vessel can be made several meters long without significant reduction in UCN
density. On the other hand, the lateral dimensions of the converter vessel
should match the size of the available neutron beam and guide it to avoid
dilution of the incident flux. The mean free path of UCN in a vessel with
such geometry is therefore only in the order of $5-10$ cm, leading to high
typical frequencies of UCN wall collisions of $50-100$ per second. It thus
becomes challenging to obtain long UCN storage time constants which however
are a prerequisite for accumulation of a high saturated UCN density. Values
measured for narrow vessels are normally well below $200$ s. For instance,
in a recent experiment on UCN production, a rather short storage time
constant of $67$ s was obtained for a vessel held at $0.7$ K, which
consisted of a $1$ m long $7\times 7$ cm$^{2}$ tube of BeO with Be windows
on each end and included a short pipe from stainless steel. That,
nonetheless, a record UCN density was obtained demonstrates the potential of
the method \cite{Zimmer/2011}. To our knowledge the Cryo-EDM collaboration
achieved with $\tau =160$ s the so far highest value for a helium converter
enclosed within matter boundaries, using a $3$ m long tubular vessel with
diameter $63$ mm, made of Be coated copper and closed off by Be windows \cite%
{Grinten/2009}.

Magnetic trapping of UCN offers a viable way for a drastic improvement of
the storage properties of the converter vessel, ultimately limited only by
the neutron lifetime $\tau _{\beta }\approx 880$ seconds. It relies on the
potential energy $\pm \mu _{\mathrm{n}}B$ of the neutron magnetic moment $%
\mu _{\mathrm{n}}\approx 60$ $\mathrm{neV/T}$ in a magnetic field $B$.
Suitable configurations of magnetic field gradients keep the low field
seeking UCN away from walls where otherwise the collisional losses occur. A
group at NIST has demonstrated storage time constants consistent with the
neutron lifetime within a helium converter equipped with a superconducting
magnetic quadrupole UCN reflector \cite{Dzhosyuk/2005,Huffman/2000}. The
apparatus was designed to perform neutron lifetime measurements, for which a
complete suppression of UCN wall contacts is mandatory. On the other hand,
for the sake of enhancing the output of a UCN source, combined magnetic and
material trapping turns out to be particularly beneficial. In addition, the
phase space for UCN accumulation can be much increased using a higher
multipole order.

In this paper we provide an analytic treatment of the rate equation for UCN
production and storage in a superfluid-helium converter confined by material
walls and surrounded by a magnetic mirror. We show that, combining a
converter vessel possessing good (but not exceptional) storage properties
with a magnetic mirror of high multipole order, one may achieve a saturated
UCN density close to the theoretical limit defined by an ideal experimental
bottle, i.e.\ a square well potential without imaginary part. That the
magnet needs to generate only part of the trapping potential is of great
practical value for constructing a device using standard superconducting
wire technology.

\section{Rate equation and system definition}

The temporal evolution of the spectral UCN density in a closed converter
irradiated with the cold beam is governed by a simple rate equation. UCN
production is characterized by a spectral rate density $p$ that depends on
the spectral flux of the incident beam, and a loss term is due to finite
lifetime $\tau $ of UCN in the converter, 
\begin{equation}
\frac{dn\left( \epsilon _{0},t\right) }{dt}=p\left( \epsilon _{0}\right) -%
\frac{n\left( \epsilon _{0},t\right) }{\tau \left( \epsilon _{0}\right) }.
\label{dn/dt}
\end{equation}%
Here we label stored neutrons by their total energy $\epsilon _{0}$, defined
as their kinetic energy at the point of lowest potential energy in the trap.
The quantities $n\left( \epsilon _{0},t\right) $ and $p\left( \epsilon
_{0}\right) $ denote the real space density, respectively, the spatial UCN
production rate density, each per energy interval of a group of UCN with
total energy in the range $\left( \epsilon _{0},\epsilon _{0}+d\epsilon
_{0}\right) $. The saturated spectral UCN density obtains when UCN losses
balance UCN production for $t\gg \tau $ after having switched on the beam.
It is given by%
\begin{equation}
n_{\infty }\left( \epsilon _{0}\right) =p\left( \epsilon _{0}\right) \tau
\left( \epsilon _{0}\right) .  \label{rho}
\end{equation}%
If one wants to fill a trap with UCN up to a cutoff energy set by the
trapping potential $V_{\mathrm{trap}}$, what matters is the saturated total
UCN density which is obtained from 
\begin{equation}
n_{\infty }=\int_{0}^{V_{\mathrm{trap}}}n_{\infty }\left( \epsilon
_{0}\right) d\epsilon _{0}.  \label{rho_sat}
\end{equation}%
Many applications of UCN sources involve filling external traps with as many
UCN as possible, followed by a long time for holding or manipulation, during
which the density in the source can be refreshed. Therefore $n_{\infty }$ is
a useful parameter of the converter to be optimized\footnote{%
For experiments involving external traps with poor storage properties it
will be better to drain UCN frequently from a partially charged source.
However, also in this case a long UCN storage time constant is an asset as
it will raise the time-averaged UCN content of the converter.}.

We consider a system as schematically shown in Fig.\ $1$. A cylindrical
converter vessel with diameter $2R$ is situated within a multipole magnet
and illuminated homogeneously by a cold neutron beam passing along the $r=0$
axis. UCN accumulation takes place over a length $L\gg R$ between a beam
window and a UCN valve fully immersed in the helium as in the apparatus
described in \cite{Golub/1983}. Shown is a butterfly valve but also
different types may be envisaged, such as an iris diaphragm valve. For
experiments at room temperature UCN are released into a window-less
extraction system with a short vertical guide section as described in \cite%
{Zimmer/2010}\footnote{%
It is also conceivable to place the UCN valve closer to (or within) the
extraction guide. This would however considerably increase the surface of
wall material exposed to the UCN during accumulation. Here we analyse the
system as shown in fig.\ $1$.}. In the section for UCN accumulation the
cylindrical wall possesses a neutron optical potential $V+iW$ \cite%
{Fermi/1936,Golub/1991,Ignatovich/1990}, with%
\begin{equation}
V=\frac{2\pi \hbar ^{2}}{m_{\mathrm{n}}}\sum_{l}N_{l}b_{l},\qquad W=\frac{%
\hbar }{2}\sum_{l}N_{l}v\sigma _{l}\left( v\right) ,  \label{V-Fermi}
\end{equation}%
where $m_{\mathrm{n}}$ is the neutron mass, $N_{l}$ is the atomic number
density of the nuclear species $l$ with coherent bound neutron scattering
length $b_{l}$, and $\sigma _{l}\left( v\right) $ is the loss cross section
(sum of cross sections for neutron capture and upscattering; $v$ is the UCN
velocity in the medium and $v\sigma _{l}\left( v\right) $ usually constant
over the whole UCN spectrum). The beam window and the UCN valve are made of
(or coated with) a material with neutron optical potential $\widetilde{V}+i%
\widetilde{W}$.

\begin{figure}[tbp]
\centering
\includegraphics[width=0.92\textwidth]{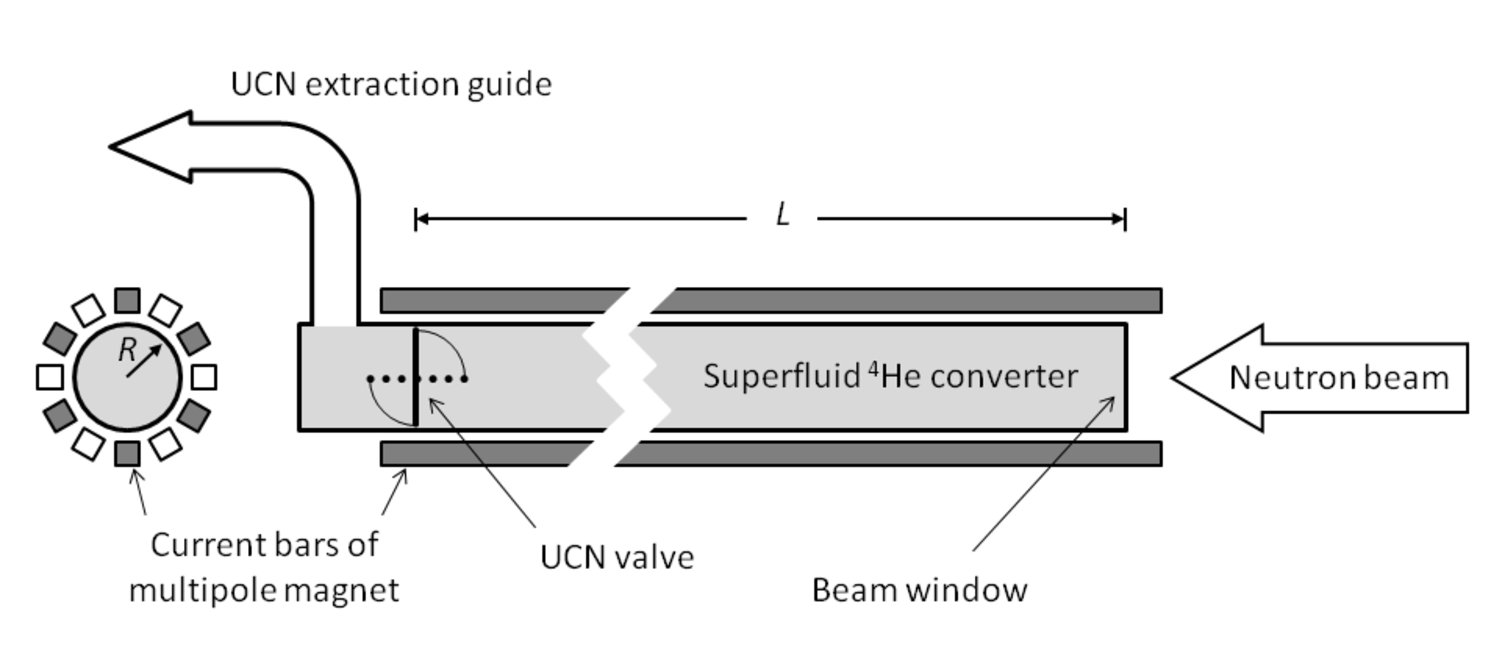}
\caption{Schematic of the UCN accumulator comprized of a superfluid $^{4}$He
converter with multipole magnet and system for UCN extraction. On the left a
cut view is shown for $n=12$; filled (open) squares indicate electric
current flowing into (out of) the plane. The neutron optical potentials are: 
$V$ for the cylindrical inner surface over the length $L$, $\widetilde{V}$
at the beam window and at the UCN valve, $\geq \widetilde{V}$ for the UCN
extraction system, and $V_{\mathrm{He}}\approx 18.5\ \mathrm{neV}$ in the
superfluid.}
\label{fig:figure1}
\end{figure}

A radial $n$-polar magnetic field with modulus%
\begin{equation}
B\left( r\right) =B_{R}\left( \frac{r}{R}\right) ^{\frac{n}{2}-1}
\label{B_R}
\end{equation}%
can be generated as shown, using a regular arrangement of an even number of $%
n$ straight current bars on the outer cylinder surface, with opposite
current in adjacent bars (in practice one employs long racetrack coils). A
neutron moving in such a field has a magnetic potential energy of%
\begin{equation}
V_{\mathrm{m}}\left( r\right) =\pm V_{\mathrm{m}R}\left( \frac{r}{R}\right)
^{\frac{n}{2}-1},\qquad V_{\mathrm{m}R}=\left\vert \mu _{\mathrm{n}%
}\right\vert B_{R},  \label{V_m}
\end{equation}%
where the upper (lower) sign in this and all subsequent equations describes
neutrons in the low (high) field seeking spin state, denoted by lfs and hfs
in the sequel. Note that we assume adiabatic spin transport and thus neglect
spin flip transitions. It is an experimentally established fact that the
validity of the adiabatic condition can be fulfilled better than needed for
our purposes, using a weak bias field in the order of a few $10\ \mathrm{mT}$
along the axis of the multipole magnet. For instance, an octupole trap has
provided UCN storage lifetimes in the order of $800\ \mathrm{s}$. Losses
occured primarily due UCN hitting a teflon piston coated with fomblin
grease, which was used for axial closure of the trap \cite{Leung/2013}. A
storage lifetime much closer to the neutron lifetime was attained in a $20$%
-pole magnetic trap, in which UCN were held without any wall collisions \cite%
{Ezhov/2005}.

Due to addition of magnetic fields to the neutron optical potentials,
trapping of UCN can become strongly spin-dependent. The trapping potential
of the converter vessel shown in Fig.\ $1$ is given by%
\begin{equation}
V_{\mathrm{trap}}=\min \left( \left( V\pm V_{\mathrm{m}R}-V_{\mathrm{He}%
}\right) \Theta \left( V\pm V_{\mathrm{m}R}-V_{\mathrm{He}}\right) ,\left( 
\widetilde{V}-V_{\mathrm{He}}\right) \Theta \left( \widetilde{V}-V_{\mathrm{%
He}}\right) \right) ,  \label{noP}
\end{equation}%
i.e.\ the minimum of the total potential (neutron optical wall potential and
magnetic interaction potential, reduced by the neutron optical potential of
the converter medium). The first argument describes the situation at the
cylindrical wall (with potential $V$), the second one at the end disks
(potential $\widetilde{V}\geq V$). The expression employs the step function $%
\Theta \left( x\right) =1$ for $x>0$ and $\Theta \left( x\right) =0$ for $%
x\leq 0$, and%
\begin{equation}
V_{\mathrm{He}}\approx 18.5\ \mathrm{neV}  \label{V-He}
\end{equation}%
is the neutron optical potential of the superfluid $^{4}$He as calculated
using eq.\ \ref{V-Fermi} with $N_{\mathrm{He}}\approx 2.18\times 10^{22}\ 
\mathrm{cm}^{-3}$ and $b_{\mathrm{He}}\approx 3.26\times 10^{-13}\ \mathrm{cm%
}$. For the lfs neutrons the lowest potential energy prevails on the line $%
r=0$, whereas for the hfs neutrons it has its minimum close to the
cylindrical wall (see Fig.\ $2$). Note however that the hfs neutrons will
only be able to leave the magnetic field if they still possess kinetic
energy at $r=0$ and hence only for $\epsilon _{0}>V_{\mathrm{m}R}$ according
to our previous definition of $\epsilon _{0}$. Since we are not further
interested in the fate of hfs neutrons unable to escape the multipolar field
after opening the UCN valve, and for simplicity of the equations to follow,
we define $\epsilon _{0}$ in the sequel commonly for both spin states as the
kinetic energy in the point of lowest magnetic field (i.e.\ on the line $r=0$%
). Equation\ \ref{rho_sat} thus describes, individually for both spin
states, UCN densities of useable UCN (i.e.\ those with $\epsilon _{0}>0$),
with $V_{\mathrm{trap}}$ as defined in eq.\ \ref{noP}. We define the
polarization of the saturated ensemble of useable UCN as%
\begin{equation}
P_{\infty }=\frac{n_{\infty ,\mathrm{lfs}}-n_{\infty ,\mathrm{hfs}}}{%
n_{\infty ,\mathrm{lfs}}+n_{\infty ,\mathrm{hfs}}}.  \label{polarization}
\end{equation}%
As shown in Fig.\ $2$ for two situations it will depend on the relative
strengths of magnetic and neutron optical potentials.

\begin{figure}[tbp]
\centering\includegraphics[width=0.92\textwidth]{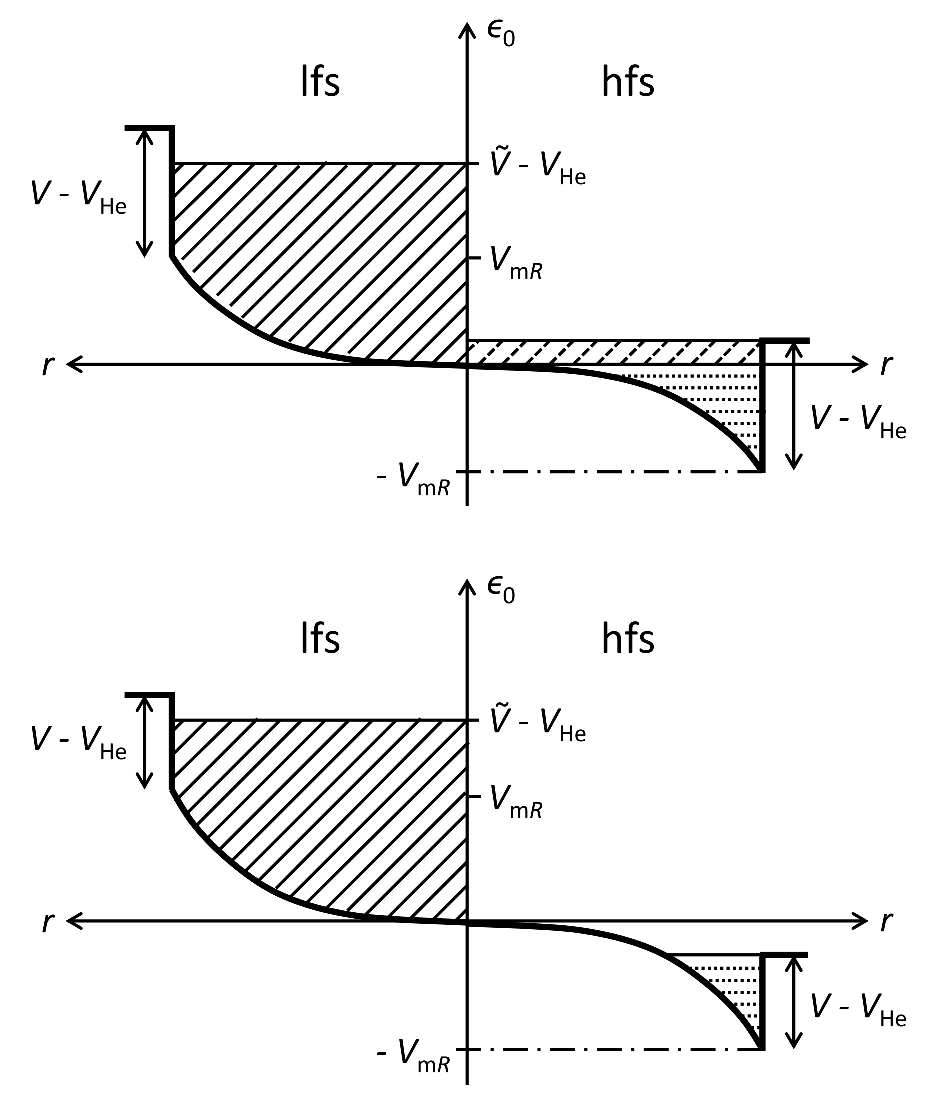}
\caption{Schematic of the magnetic and neutron optical potentials in the
closed UCN accumulator shown in Fig. $1$. In the upper figure, $V-V_{\mathrm{%
He}}>V_{\mathrm{m}R}$. Low field seeking (lfs) UCN with total energy $%
\protect\epsilon _{0}<\widetilde{V}-V_{\mathrm{He}}$ are trapped (solid line
shade). High field seeking (hfs) UCN with $0<\protect\epsilon _{0}<V-V_{%
\mathrm{He}}-V_{\mathrm{m}R}$ are trapped, too (dashed-line shade), leading
to polarization $P_{\infty }<1$ according to eq.\ \protect\ref{polarization}%
. The lower figure illustrates the situation $V-V_{\mathrm{He}}<V_{\mathrm{m}%
R}$ where hfs UCN with $\protect\epsilon _{0}>0$ are free to escape, hence $%
P_{\infty }=1$. Only hfs UCN with $\protect\epsilon _{0}<0$ (horizontal
dotted shade) are trapped which however will stay in the regions with strong
magnetic field when opening the UCN valve.}
\label{fig:figure2}
\end{figure}

\section{UCN losses from the converter}

The inverse of the time constant appearing in the loss term in eq.\ \ref%
{dn/dt} is comprized of several contributions,%
\begin{equation}
\tau ^{-1}\left( \epsilon _{0},T\right) =\tau _{\mathrm{wall}}^{-1}\left(
\epsilon _{0}\right) +\tau _{\mathrm{slit}}^{-1}\left( \epsilon _{0}\right)
+\tau _{\mathrm{up}}^{-1}\left( T\right) +\tau _{\mathrm{abs}}^{-1}+\tau _{%
\mathrm{depol}}^{-1}\left( \epsilon _{0}\right) +\tau _{\beta }^{-1},
\label{tau-rate}
\end{equation}%
where the argument indicates now also the dependence on the temperature $T$
of the converter. From left to right, they describe UCN loss at collisions
with the walls of the converter vessel, escape of UCN through an imperfectly
closed UCN valve and through slits caused by manufacturing imperfections of
the vessel assembly, upscattering by thermal excitations in the helium,
absorption by $^{3}$He impurities, UCN depolarization at wall collisions,
and neutron beta decay. Note that, using an unpolarized beam for neutron
conversion, the rate constant $\tau _{\mathrm{depol}}^{-1}$ may become
relevant only if trapping is at least partly magnetic. In an experimental
study a depolarization probability per wall collision of $7\times 10^{-6}$
was measured for a bottle made of Be \cite{Serebrov/2000}. Hence, $\tau _{%
\mathrm{depol}}^{-1}<\tau _{\beta }^{-1}$ even for the fastest neutrons in
the narrow trap geometry discussed here. Like the first two rate constants
in eq.\ \ref{tau-rate}, $\tau _{\mathrm{depol}}^{-1}$ will be further
suppressed due to the multipole magnet (provided that spin transport is
adiabatic), and we therefore neglect it. For temperatures below $1$ K \cite%
{Golub/1983,Golub/1979},%
\begin{equation}
\tau _{\mathrm{up}}^{-1}\left( T\right) \approx \frac{\left( T\left[ \mathrm{%
K}\right] \right) ^{7}}{100\,\mathrm{s}},  \label{tau_up}
\end{equation}%
so that for $T<0.5$ K, $\tau _{\mathrm{up}}^{-1}$ contributes with less than 
$10\%$ of $\tau _{\beta }^{-1}$. The rate constant $\tau _{\mathrm{abs}%
}^{-1} $ can be suppressed below any relevant level by purification of the
helium from $^{3}$He using a superleak \cite%
{Zimmer/2010,Yoshiki/2005,Yoshiki/1994} or the heat flush technique \cite%
{McClintock/1978}. As a result, we are left with $\tau _{\mathrm{wall}}^{-1}$
and $\tau _{\mathrm{slit}}^{-1}$ as dominating contributions, and the rate
constant due to neutron decay sets an ultimate lower limit for a perfect
converter.

For the losses due to wall collisions we want to apply an analytic
description. If we assume the trapped UCN in mechanical equilibrium we can
use formulas derived in the book \cite{Golub/1991} where the authors
analyzed the effect of the earth's gravitational field on neutrons moving in
a bottle. We adapt the notation to our case and replace the height parameter 
$h$ by the radial coordinate $r$ characterizing the multipolar magnetic
field strength. We neglect the gravitational field, which is a good
approximation for a horizontal source with less than $10$ cm diameter. The
kinetic neutron energy for the two spin states is then given by $E\left(
r\right) =\epsilon _{0}\mp \left\vert V_{\mathrm{m}}\left( r\right)
\right\vert $. The energy,%
\begin{equation}
E:=E\left( R\right) =\epsilon _{0}\mp V_{\mathrm{m}R},  \label{E}
\end{equation}%
is positive for the extractable hfs neutrons which can always explore the
whole trap. The low field seeking neutrons on the other hand may have too
low energy to hit the walls. For $E>0$ eq.\ \ref{E} defines the energy of
neutron impact. Employing a formula given in ref.\ \cite{Golub/1991}, the
angular averaged probability $\overline{\mu }\left( E\right) $ for UCN loss
during a collision with the cylindrical wall of the helium filled converter
vessel can be written as

\begin{equation}
\overline{\mu }\left( E\right) =2f\mbox{ Re}\left( \frac{V^{\prime }}{E}%
\arcsin \sqrt{\frac{E}{V^{\prime }}}-\sqrt{\frac{V^{\prime }}{E}-1}\right)
\qquad \text{for }E\leq V^{\prime },  \label{mu}
\end{equation}%
valid for a neutron optical potential $V+iW$ with small losses, i.e.$\
f=W/V\ll 1$, and writing%
\begin{equation}
V^{\prime }=V-V_{\mathrm{He}},
\end{equation}%
with $V_{\mathrm{He}}$ given in eq.\ \ref{V-He}. For convenience in later
calculations we have included a projection onto the real part of the
expression. It offers a handy formulation of the case where neutrons have
too low energy to hit the wall, ensuring that $\overline{\mu }\left(
E<0\right) =0$ without need to specify the range of $E$ in advance as
positive. The function $\overline{\mu }\left( E\right) $ rises monotonously
with $E$ from $\overline{\mu }=0$ for $E=0$ to $\overline{\mu }=\pi f$ for $%
E=V^{\prime }$. For $E>V^{\prime }$ we may set $\overline{\mu }=1$ since we
are not interested here in calculating the dynamics of marginally trapped
neutrons. For $E=V^{\prime }/2$, $\overline{\mu }\approx 1.14f$. Note that,
since we consider a long trap for which $2\pi RL\gg \pi R^{2}$, we will
neglect losses due to $\widetilde{W}$ at the end disks. The magnetic
multipole suppresses wall losses of lfs UCN for several reasons. First, only
a fraction of them has sufficient energy to hit the lossy wall. Second,
those lfs neutrons with $\epsilon _{0}>V_{\mathrm{m}R}$ hit the wall due to
magnetic deceleration with a reduced kinetic energy $E$ (eq.\ \ref{E}),
leading to reduced losses due to $\overline{\mu }\left( \epsilon _{0}-V_{%
\mathrm{m}R}\right) <\overline{\mu }\left( \epsilon _{0}\right) $. Third,
the rate of wall collisions of these neutrons is reduced as well, leading to
a further suppression in the expressions for $\tau _{\mathrm{wall}}^{-1}$
and $\tau _{\mathrm{slit}}^{-1}$, which we calculate next.

In mechanical equilibrium, a group of neutrons with total energy in the
range $\left( \epsilon _{0},\epsilon _{0}+d\epsilon _{0}\right) $ will
occupy the accessible phase space in the trap with uniform density. As a
result of phase space transformation under the influence of a conservative
potential (see section 4.3.1 in the book \cite{Golub/1991}), the real space
spectral UCN densities in different positions are related by 
\begin{equation}
n\left( \epsilon _{0},t,r\right) =\mbox{ Re}\sqrt{\frac{\epsilon _{0}\mp
\left\vert V_{\mathrm{m}}\left( r\right) \right\vert }{\epsilon _{0}}}%
n\left( \epsilon _{0},t,0\right) ,  \label{n_r}
\end{equation}%
where projection onto the real part ensures $n\left( \epsilon
_{0},t,r\right) =0$ for the lfs UCN for $r>R^{\ast }$ defined by $\epsilon
_{0}=\left\vert V_{\mathrm{m}}\left( R^{\ast }\right) \right\vert $. We
define an effective volume of the source for neutrons with total energy $%
\epsilon _{0}$ as%
\begin{equation}
\gamma \left( \epsilon _{0}\right) =2\pi L\mbox{ Re}\int_{0}^{R}\sqrt{\frac{%
\epsilon _{0}\mp \left\vert V_{\mathrm{m}}\left( r\right) \right\vert }{%
\epsilon _{0}}}rdr.
\end{equation}%
The spectral UCN density averaged over the entire volume of the converter is
then given by%
\begin{equation}
n\left( \epsilon _{0},t\right) =\gamma ^{\prime }\left( \epsilon _{0}\right)
n\left( \epsilon _{0},t,0\right) ,  \label{Int-n}
\end{equation}%
and the reduced quantity,%
\begin{equation}
\gamma ^{\prime }\left( \epsilon _{0}\right) =\frac{\gamma \left( \epsilon
_{0}\right) }{\pi R^{2}L}=2\mbox{ Re}\int_{0}^{1}\sqrt{1\mp \frac{V_{\mathrm{%
m}R}}{\epsilon _{0}}r^{\frac{n}{2}-1}}rdr,  \label{gamma'}
\end{equation}%
was derived using eq.\ \ref{V_m}. Values for $\gamma ^{\prime }$ listed in
Table $1$ show that, the higher the multipolarity, the less significant
becomes the reduction of the density of the lfs neutrons with respect to a
square well potential of same depth.%
%TCIMACRO{\TeXButton{B}{\begin{table}[tbp] \centering}}%
%BeginExpansion
\begin{table}[tbp] \centering%
%EndExpansion
$%
\begin{tabular}{l|lllllll}
$n\left\backslash \frac{\epsilon _{0}}{V_{\mathrm{m}R}}\right. $ & $\frac{1}{%
10}$ & $\frac{1}{5}$ & $\frac{1}{2}$ & $1$ & $\frac{4}{3}$ & $\frac{8}{5}$ & 
$2$ \\ \hline
$4$ & $0.\,\allowbreak 005$ & $0.021$ & $0.133$ & $0.533$ & $0.696$ & $0.758$
& $0.813$ \\ 
$6$ & $0.\,\allowbreak 067$ & $0.133$ & $0.333$ & $0.667$ & $0.778$ & $0.822$
& $0.862$ \\ 
$8$ & $0.159$ & $0.252$ & $0.465$ & $0.739$ & $0.824$ & $0.859$ & $0.89$ \\ 
$10$ & $0.248$ & $0.351$ & $0.555$ & $0.785$ & $0.855$ & $0.883$ & $0.909$
\\ 
$12$ & $0.325$ & $0.429$ & $0.619$ & $0.818$ & $0.876$ & $0.9$ & $0.922$ \\ 
$14$ & $0.39$ & $0.492$ & $0.667$ & $0.841$ & $0.892$ & $0.913$ & $0.932$%
\end{tabular}%
$%
\caption{Values for $\gamma
^{\prime }$ for low field seeking neutrons, as defined with the upper sign in eq.\ \ref{gamma'} for various values of $\epsilon _{0}/V_{\mathrm{m}R}$.}%
\label{TableKey}%
%TCIMACRO{\TeXButton{E}{\end{table}} }%
%BeginExpansion
\end{table}
%EndExpansion
The spectral current density of neutrons at any point in the vessel, per
unit area and per energy interval about $\epsilon _{0}$, is given by the gas
kinetic relation%
\begin{equation}
J\left( \epsilon _{0},t,r\right) =\frac{1}{4}n\left( \epsilon
_{0},t,r\right) v\left( \epsilon _{0},r\right) .  \label{J}
\end{equation}%
The spectral rate of UCN collisions with the cylindrical wall of the helium
container is given by $2\pi RLJ\left( \epsilon _{0},t,R\right) $. The speed $%
v\left( \epsilon _{0},R\right) $ of the neutrons as they hit the wall is
related to the speed at $r=0$ through%
\begin{equation}
v\left( \epsilon _{0},R\right) =\mbox{ Re}\sqrt{\frac{\epsilon _{0}\mp V_{%
\mathrm{m}R}}{\epsilon _{0}}}v\left( \epsilon _{0},0\right) .  \label{v}
\end{equation}%
With eq.\ \ref{n_r} we obtain%
\begin{equation}
J\left( \epsilon _{0},t,R\right) =\frac{\epsilon _{0}\mp V_{\mathrm{m}R}}{%
\epsilon _{0}}J\left( \epsilon _{0},t,0\right) \Theta \left( \epsilon
_{0}\mp V_{\mathrm{m}R}\right) ,  \label{J-R}
\end{equation}%
with the step function $\Theta \left( x\right) $ as already used in eq.\ \ref%
{noP}. For the loss term in eq.\ \ref{dn/dt} due to collisions with the
cylindrical wall we can thus write%
\begin{equation}
\frac{n\left( \epsilon _{0},t\right) }{\tau _{\mathrm{wall}}\left( \epsilon
_{0}\right) }=\frac{2}{R}\overline{\mu }\left( \epsilon _{0}\mp V_{\mathrm{m}%
R}\right) J\left( \epsilon _{0},t,R\right) .
\end{equation}%
Further evaluation using eq.\ \ref{J-R}, eq.\ \ref{J} for $r=0$, eq.\ \ref%
{Int-n}, and inserting $\overline{\mu }$ from eq.\ \ref{mu} leads us to%
\begin{equation}
\tau _{\mathrm{wall}}^{-1}\left( \epsilon _{0}\right) =\frac{v\left(
\epsilon _{0},0\right) }{R\gamma ^{\prime }\left( \epsilon _{0}\right) }f%
\frac{V^{\prime }}{\epsilon _{0}}\mbox{ Re}\left( \arcsin \sqrt{\frac{%
\epsilon _{0}\mp V_{\mathrm{m}R}}{V^{\prime }}}-\sqrt{\frac{\epsilon _{0}\mp
V_{\mathrm{m}R}}{V^{\prime }}\left( 1-\frac{\epsilon _{0}\mp V_{\mathrm{m}R}%
}{V^{\prime }}\right) }\right) .  \label{tau_wall}
\end{equation}

For the calculation of the corresponding expression for losses through slits
it is reasonable to assume them to be situated at $r=R$, e.g.\ at the seam
of the tube or at its connections to the circular windows for the cold beam.
Assuming that any UCN hitting a slit will be lost and denoting the total
surface of all slits by $A$, their contribution to the loss term in
eq.\noindent\ \ref{dn/dt} is given by%
\begin{equation}
\frac{n\left( \epsilon _{0},t\right) }{\tau _{\mathrm{slit}}\left( \epsilon
_{0}\right) }=\frac{2}{R}\frac{A}{2\pi RL}J\left( \epsilon _{0},t,R\right) ,
\end{equation}%
neglecting the small surface of the disks at the ends. Hence,%
\begin{equation}
\tau _{\mathrm{slit}}^{-1}\left( \epsilon _{0}\right) =\frac{Av\left(
\epsilon _{0},0\right) }{4\gamma \left( \epsilon _{0}\right) }\frac{\epsilon
_{0}\mp V_{\mathrm{m}R}}{\epsilon _{0}}\Theta \left( \epsilon _{0}\mp V_{%
\mathrm{m}R}\right) ,
\end{equation}%
and we see that, for the lfs neutrons, the ordinary gas kinetic expression
represented in the first fraction on the right side becomes reduced for $V_{%
\mathrm{m}R}>0$ for the same reason as the wall losses discussed before.

\section{UCN production}

We first consider UCN production in absence of the magnetic multipole field.
For homogeneous irradiation with the cold neutron beam guided through the
converter, neglecting decrease of intensity due to reflection losses and
neutron scattering in the helium, the UCN production rate density is
position independent and given by 
\begin{equation}
p_{0}=\int_{0}^{V_{\mathrm{trap}}}p_{0}\left( \epsilon _{0}\right) d\epsilon
_{0}=KV_{\mathrm{trap}}^{3/2}.  \label{p_0}
\end{equation}%
The $V_{\mathrm{trap}}^{3/2}$ dependence follows for a homogeneous
population of states within a sphere in momentum space with spectral UCN
production rate density%
\begin{equation}
p_{0}\left( \epsilon _{0}\right) =\frac{3}{2}K\sqrt{\epsilon _{0}}.
\label{p_0-diff}
\end{equation}%
The factor $K$ due to single phonon emission has been calculated on the
basis of neutron scattering data and confirmed in several experiments \cite%
{Schmidt-Wellenburg/2009,Zimmer/2007,Baker/2003,Korobkina/2002,Masuda/2002,Ageron/1978,Golub/1977}%
, albeit with modest experimental accuracy limited by detection efficiency
and other corrections. For UCN with maximum energy determined by $V_{\mathrm{%
trap}}=V-V_{\mathrm{He}}\approx 233\ \mathrm{neV}$ for Be or Ni with natural
isotopic composition, it is given by 
\begin{equation}
K\approx 5\mathrm{s}^{-1}\mathrm{cm}^{-3}\Phi _{0.89\mathrm{nm}}\left[ 10^{9}%
\mathrm{cm}^{-2}\mathrm{s}^{-1}\mathrm{nm}^{-1}\right] /\left( 233\mathrm{neV%
}\right) ^{3/2},
\end{equation}%
where $\Phi _{0.89\mathrm{nm}}$ is the differential unpolarized neutron flux
density at a neutron wavelength of $0.89$ nm. The flux unit is chosen
numerically close to values available at existing facilities, e.g.\ the
monochromatic beam H172A at the ILL \cite{Piegsa/2014}. An additional,
usually smaller contribution to UCN production is due to multi-phonon
processes.

When adding the multipolar magnetic field, the spectral UCN production rate
density becomes dependent on position and spin state,%
\begin{equation}
p_{\mathrm{lfs(hfs)}}\left( \epsilon _{0},r\right) =\frac{3}{4}K\left(
r\right) \func{Re}\sqrt{\epsilon _{0}\mp \left\vert V_{m}\left( r\right)
\right\vert },
\end{equation}%
where the $r$ dependence of $K$ accounts for a spatially varying flux
density of the neutron beam. The factor $1/2$ with respect to eq.\ \ref%
{p_0-diff} holds for an unpolarized beam incident on the converter, as
always assumed hereafter. For homogeneous irradiation, $K\left( r\right) =K$%
, and the spatially averaged spectral UCN production rate density for the
two spin states can be expressed in terms of the normalized effective volume
from eq.\ \ref{gamma'}, i.e.%
\begin{equation}
p_{\mathrm{lfs(hfs)}}\left( \epsilon _{0}\right) =\frac{3}{4}K\gamma
^{\prime }\left( \epsilon _{0}\right) \sqrt{\epsilon _{0}}.  \label{p_diff}
\end{equation}%
Without magnetic field, $p_{0,\mathrm{lfs}}=p_{0,\mathrm{hfs}}=p_{0}/2$ with 
$p_{0}$ given in eq.\ \ref{p_0}, whereas with field,%
\begin{equation}
p_{\mathrm{lfs(hfs)}}=\int_{0}^{V_{\mathrm{trap}}}p_{\mathrm{lfs(hfs)}%
}\left( \epsilon _{0}\right) d\epsilon _{0}
\end{equation}%
are different due to the spin-dependent $\gamma ^{\prime }\left( \epsilon
_{0}\right) $ and $V_{\mathrm{trap}}$ from eq.\ \ref{noP}. We note
particularly that the ratio of total production rates for lfs UCN with the
magnetic multipole switched on and switched off,%
\begin{equation}
\kappa =\frac{p_{\mathrm{lfs}}}{p_{0,\mathrm{lfs}}},  \label{kappa}
\end{equation}%
is smaller than unity due to the phase space reduction by the magnetic
multipole. The values quoted in Table $2$ demonstrate a positive effect of
high multipolar order $n$ on $\kappa $ and hence on the saturated UCN
density calculated in the next section. There are however practical limits.
First, thermal insulation between the magnet and the much colder helium
container necessitates an annular gap over which the field would drop too
strongly if $n$ is chosen too large. Second, the maximum field strength
achievable with a given maximum current density in the current bars around
the converter of given diameter decreases with $n$. For $R=5$ cm, and taking
into account the results given in the next section, $n\approx 12$ turns out
to be a reasonable choice. 
%TCIMACRO{\TeXButton{B}{\begin{table}[tbp] \centering}}%
%BeginExpansion
\begin{table}[tbp] \centering%
%EndExpansion
$%
\begin{tabular}{l|llll}
$n\left\backslash \frac{V_{\mathrm{m}R}}{V_{\mathrm{trap}}}\right. $ & $1$ & 
$\frac{3}{4}$ & $\frac{5}{8}$ & $\frac{1}{2}$ \\ \hline
$4$ & $0.229$ & $0.37$ & $0.456$ & $0.551$ \\ 
$6$ & $0.4$ & $0.517$ & $0.585$ & $0.659$ \\ 
$8$ & $0.512$ & $0.609$ & $0.665$ & $0.725$ \\ 
$10$ & $0.589$ & $0.672$ & $0.72$ & $0.77$ \\ 
$12$ & $0.645$ & $0.718$ & $0.759$ & $0.802$ \\ 
$14$ & $0.688$ & $0.752$ & $0.788$ & $0.827$%
\end{tabular}%
$%
\caption{Values for $\kappa$ as defined in eq.\ \ref{kappa} for various values of $n$ and
$V_{\mathrm{m}R}/V_{\mathrm{trap}}$.}\label{TableKey copy(1)}%
%TCIMACRO{\TeXButton{E}{\end{table}}}%
%BeginExpansion
\end{table}%
%EndExpansion

\section{Saturated UCN density}

The spatially averaged saturated spectral densities for lfs and hfs UCN
follow from eq.\ \ref{rho} with eq.\ \ref{p_diff}, i.e.%
\begin{equation}
n_{\infty ,\mathrm{lfs(hfs)}}\left( \epsilon _{0}\right) =\frac{3}{4}K\gamma
^{\prime }\left( \epsilon _{0}\right) \sqrt{\epsilon _{0}}\tau \left(
\epsilon _{0}\right) ,
\end{equation}%
using the corresponding sign in eqs.\ \ref{gamma'} and in the expression for 
$\tau \left( \epsilon _{0}\right) $. Hence, using eq.\ \ref{rho_sat}\ and
writing out all arguments relevant for characterizing the multipolar
magnetic field and the converter, the saturated total mean UCN densities in
the converter are given by%
\begin{equation}
n_{\infty ,\mathrm{lfs(hfs)}}\left( R,V_{\mathrm{m}R},n,V,f,\widetilde{V}%
,T\right) =\frac{3}{4}K\int_{0}^{V_{\mathrm{trap}}}\frac{\gamma ^{\prime
}\left( \epsilon _{0},V_{\mathrm{m}R},n\right) \sqrt{\epsilon _{0}}}{\tau
^{-1}\left( \epsilon _{0},R,V_{\mathrm{m}R},n,V,f,T\right) }d\epsilon _{0}.
\label{n_inf}
\end{equation}%
The dependence on $\widetilde{V}$ is contained in the upper limit of
integration, see eq.\ \ref{noP}. From the various contributions to the rate
constant $\tau ^{-1}$ (see eq.\ \ref{tau-rate}) we retain the terms due to
wall collisions, upscattering (eq.\ \ref{tau_up}) and neutron beta decay,
assuming that the wall losses can entirely be described by eq.\ \ref%
{tau_wall} and that there is no $^{3}$He in the converter and no slit in the
vessel.

\begin{figure}[tbp]
\centering\includegraphics[width=1.0\textwidth]{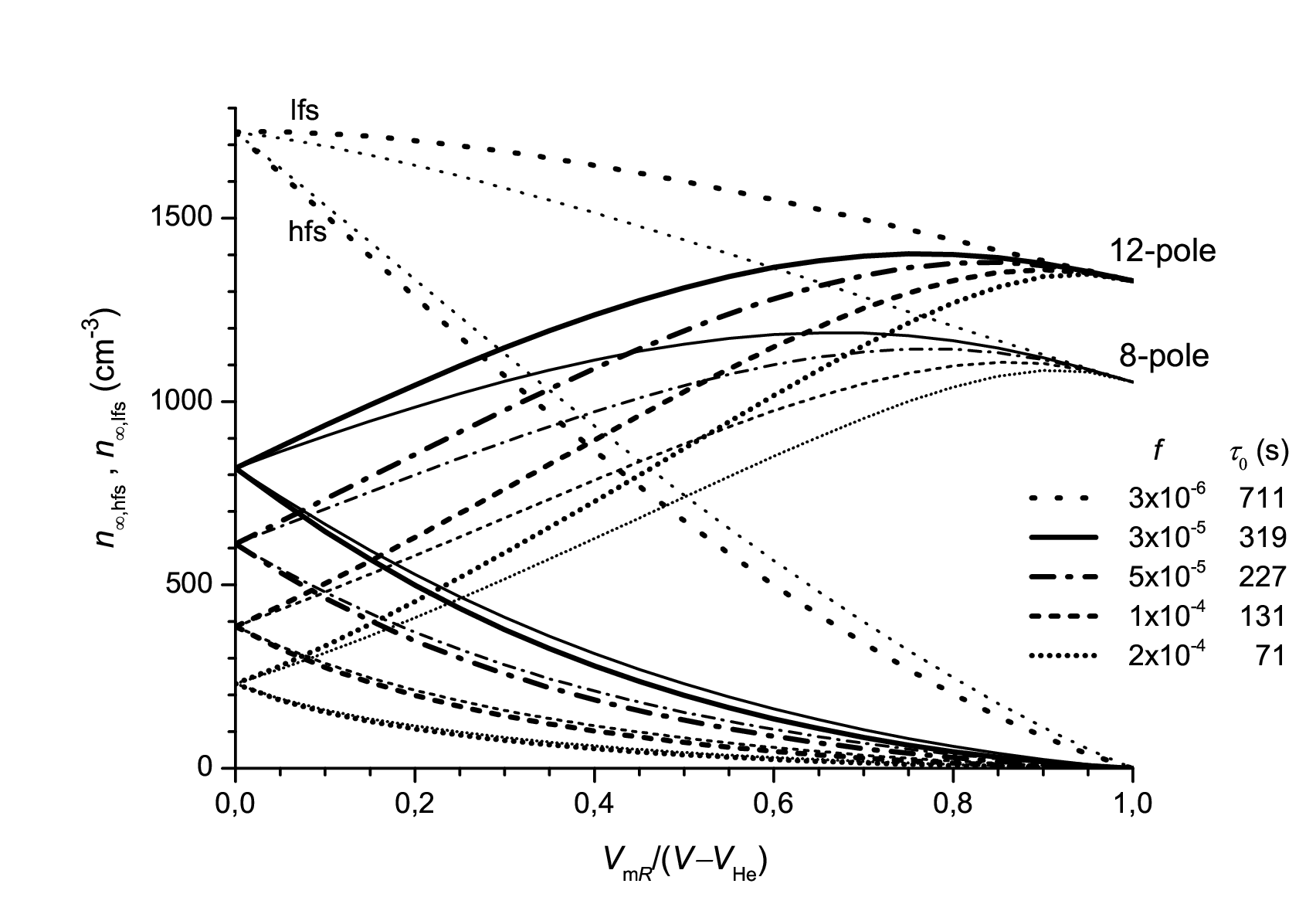}
\caption{Saturated densities of low field and high field seeking UCN in a
converter vessel with diameter $10$ cm, held at $0.5$ K and surrounded by a $%
12$-pole, respectively, $8$-pole magnet. Calculations employ eq.\ \protect
\ref{n_inf} for $\widetilde{V}=V=252$ neV (e.g.\ for a converter vessel made
entirely of Be), and for various values of $f=W/V$. The solid lines show
results for the best value of $f$ previously achieved for Be, while the
upmost curves show the situation for unrealistically low $f$ for
illustration. Values are given for an unpolarized differential neutron flux
density of $\Phi _{0.89\mathrm{nm}}=10^{9}\ $cm$^{-2}$s$^{-1}$nm$^{-1}$ at $%
\protect\lambda =0.89$ nm, as available at the monochromatic beam position
H172A at the ILL. A characteristic time constant $\protect\tau _{0}$ is
calculated for neutrons with velocity $v=\frac{3}{4}v_{\max }$, including in
eq.\ \protect\ref{tau-rate} the rates $\protect\tau _{\protect\beta }^{-1}$, 
$\protect\tau _{\mathrm{up}}^{-1}$ and the wall collisional losses
calculated using eq.\ \protect\ref{tau_wall} for $V_{\mathrm{m}R}=0$ and $%
\protect\epsilon _{0}=\frac{9}{16}V_{\mathrm{trap}}$.}
\label{fig:figure3}
\end{figure}

A first calculation of $n_{\infty ,\mathrm{lfs}}$ and $n_{\infty ,\mathrm{hfs%
}}$ was performed for a vessel featuring the neutron optical potential of Be
at all walls (i.e.$\ \widetilde{V}=V=252$\ $\mathrm{neV}$). Beryllium has
become a standard material for UCN trapping, with a best reported
experimental value of $f=3\times 10^{-5}$ in the low temperature limit \cite%
{Brys/2005,Alfimenkov/1992}, despite a much smaller theoretical value (the
finding that this was never reached was termed "anomalous losses" and has
triggered many experimental investigations and speculations). However, it
might be more realistic to consider also worse values for $f$, assuming that
efficient cleaning procedures cannot be applied in situ (e.g.\ baking is
excluded in presence of indium seals). Figure $3$ shows results exemplary
for multipole order $n=8$ and $n=12$, as a function of the magnetic trapping
potential at the cylindrical wall of the vessel, normalized to the trapping
depth without magnetic field. The density of lfs UCN increases with $n$ as
expected due to the increase in trapping phase space, while that of the hfs
neutrons decreases. Hence, for partial magnetic trapping of the lfs neutrons
(characterized by $V_{\mathrm{m}R}<V-V_{\mathrm{He}}$, see Fig.\ $2$),
higher multipole order leads to higher UCN polarization defined in eq.\ \ref%
{polarization}. For instance, for a $12$-pole with $B_{\mathrm{R}}=2.5\ 
\mathrm{T}$, one obtains $P_{\infty }=86\%$ for $f=3\times 10^{-5}$ while
for a worse $f=2\times 10^{-4}$ it improves to $P_{\infty }=97\%$. As also
obvious from the curves, the poorer the neutron optical UCN storage
performance, the larger will be the improvement of lfs UCN density due to
the multipole magnet. For the experimental cases reported in the
introduction, with measured loss rate ratios as high as $\tau ^{-1}/\tau
_{\beta }^{-1}\simeq 5.5$ \cite{Grinten/2009}, respectively $13$ \cite%
{Zimmer/2011}, a multipole magnet would indeed be very useful. Some of the
curves for $n_{\infty ,\mathrm{lfs}}$ exhibit a maximum for values $V_{%
\mathrm{m}R}/V_{\mathrm{trap}}<1$. This can be understood as resulting from
the competition of storage time constant $\tau $ and the effective trap
volume $\gamma ^{\prime }$ entering eq.\ \ref{n_inf}. For bad values of $f$
the optimum obtains for $V_{\mathrm{m}R}/V_{\mathrm{trap}}$ close to $1$,
while for a trap with excellent storage properties the multipole field
reduces the UCN density even at low field values because the factor $\gamma
^{\prime }<1$ then dominates over a marginal gain in $\tau $. For
illustration of this behaviour we also added a curve for an unrealistic
converter vessel with hypothetical $f=3\times 10^{-6}$.

\begin{figure}[tbp]
\centering\includegraphics[width=1.0\textwidth]{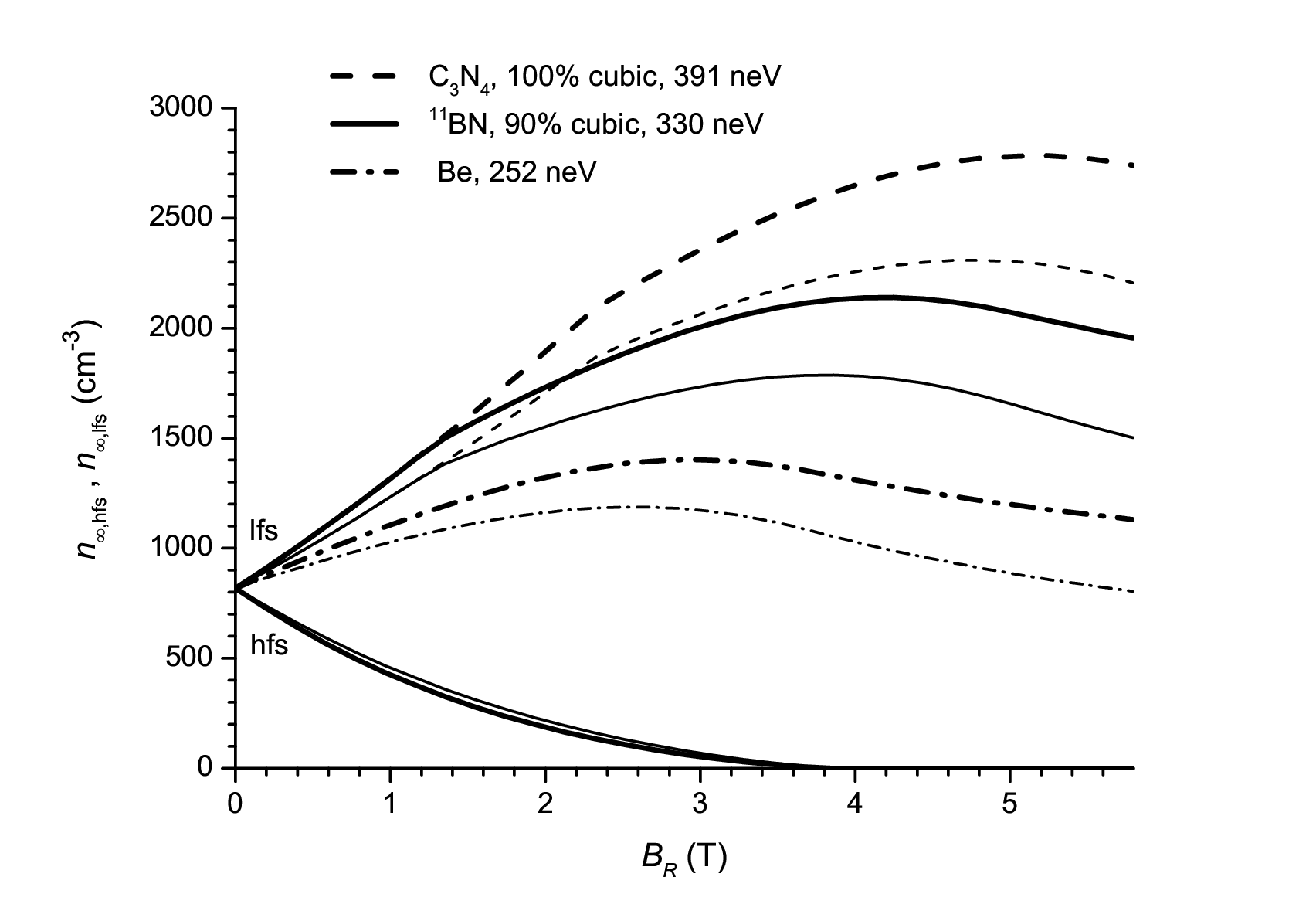}
\caption{Saturated densities of low field and high field seeking UCN in
converter vessels with diameter $10$ cm, $V=252$ neV, $f=3\times 10^{-5}$, $%
T=0.5$ K, and beam window and UCN valve (see Fig.$\ 1$) coated with three
different materials with values of $\widetilde{V}$ as indicated in the
legend. The thicker of each pair of curves is for multipolarity $n=12$, the
thinner for $n=8$, and all values are given for $\Phi _{0.89\mathrm{nm}%
}=10^{9}\ $cm$^{-2}$s$^{-1}$nm$^{-1}$ (unpolarized). The kinks visible for
the two upper pairs of curves appear at field values corresponding to $V_{%
\mathrm{m}R}=\widetilde{V}-V$. Densities of the hfs UCN for $\widetilde{V}%
\geq V$ are independent on $\widetilde{V}$ but are smaller for the higher
multipole order.}
\label{fig:figure4}
\end{figure}

Next we consider an interesting further opportunity for buildup of a high
lfs UCN density which takes advantage of the fact that the multipole magnet
increases not only storage time constants but also the potential energy of
the neutron at the cylindrical wall. As a result, the trapping depth of the
converter vessel becomes larger if the disks providing axial confinement are
made of a material with larger neutron optical potential $\widetilde{V}>V$
(remember eq.\ \ref{noP} and see Fig.\ $1$). Since the surface of the disks
is small, one may even employ materials which would be unsuitable for the
entire vessel, for an unfavorably large $\widetilde{f}=\widetilde{W}/%
\widetilde{V}$ or because coating the tubular section with sufficient
quality might be unavailable. While diamondlike carbon has already been
studied in some detail \cite%
{Atchison/2007,Atchison/2006,Atchison/2006b,Grinten/1999}, further candidate
materials able to extend the spectrum for UCN trapping beyond the Be cutoff
have been the scope of recent investigations \cite{Sobolev/2010}.
Particularly promising is boron nitride in the cubic phase (cBN). Its
neutron optical potential of $324$\ $\mathrm{neV}$ is even larger than that
of diamond ($304$\ $\mathrm{neV}$) but due to the large absorption cross
section of the isotope $^{10}$B, $\widetilde{f}=1.5\times 10^{-2}$ is also
excessively large. Enrichment of the weakly absorbing $^{11}$B however may
reduce $\widetilde{f}$ down to $3.3\times 10^{-5}$, along with a further
increase of $\widetilde{V}$ to a theoretical value of $351$\ $\mathrm{neV}$.
Using experiments on transmission with time-of-flight analysis and cold
neutron reflectometry, the authors of ref.\ \cite{Sobolev/2010} have
demonstrated a value of $305\pm 15$\ $\mathrm{neV}$ for their $2$ $\mathrm{%
\mu m}$ thick deposit of cBN (with natural isotopic composition) on a
circular silicon waver. The deviation from the ideal value is due to a cubic
phase content of $90\%$, which was measured independently by IR
spectroscopy. For highly enriched material, and assuming the same cubic
phase content, one may expect a neutron optical potential of about $330$\ $%
\mathrm{neV}$.

\begin{figure}[tbp]
\centering\includegraphics[width=1.0\textwidth]{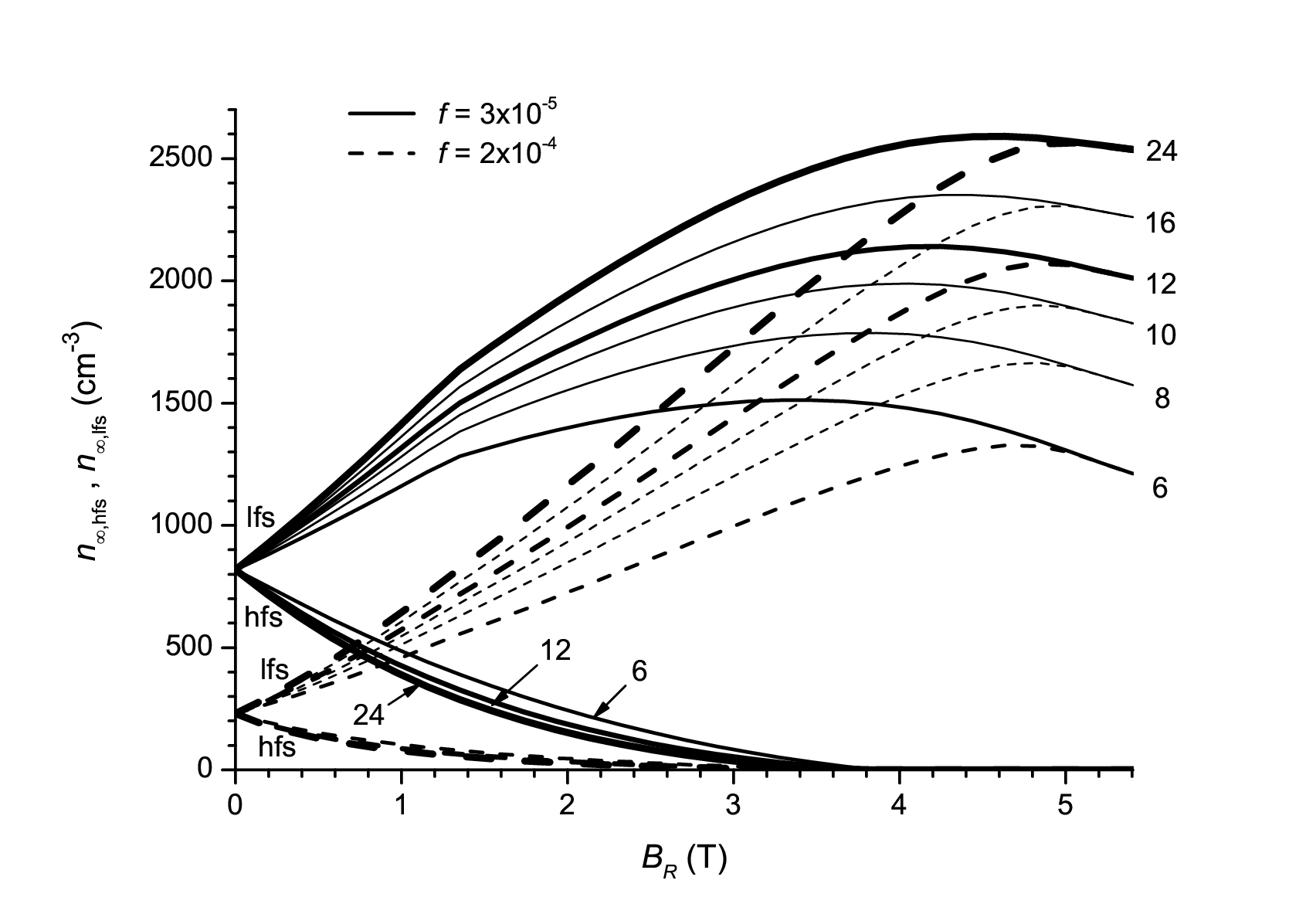}
\caption{Saturated densities of low field and high field seeking UCN in
converter vessels with diameter $10$ cm, $V=252$ neV, $T=0.5$ K, $f=3\times
10^{-5}$ (solid curves) or $f=2\times 10^{-4}$ (dashed). Beam window and UCN
valve (see Fig.$\ 1$) are coated with c$^{11}$BN with $\widetilde{V}=330$
neV. The magnetic multipole order is varied between $n=6$ and $24$. Values
are given for $\Phi _{0.89\mathrm{nm}}=10^{9}\ $cm$^{-2}$s$^{-1}$nm$^{-1}$
(unpolarized).}
\label{fig:figure5}
\end{figure}

Figure $4$ shows saturated UCN densities calculated for a trap with a
Be-coated cylindrical wall with $f=3\times 10^{-5}$ and with the axial UCN
confinement provided by Be, $^{11}$BN ($90\%$ cubic), or cubic C$_{3}$N$_{4}$%
. The latter features an extraordinarily large theoretical value of $%
\widetilde{V}=391$\ $\mathrm{neV}$. The curves for $\widetilde{V}>V$ for the
lfs UCN start with a slope larger than in the case $\widetilde{V}=V$
(dash-dotted). This is due to a $B_{R}$ dependent increase of the
integration range, $V_{\mathrm{trap}}$, in eq.\ \ref{n_inf}, as long as $V_{%
\mathrm{m}R}<\widetilde{V}-V$. Kinks in the curves appear at magnetic field
values corresponding to $V_{\mathrm{m}R}=\widetilde{V}-V$ where the full
trapping depth is reached. While for a high quality Be trap with $\widetilde{%
V}=V=252$\ $\mathrm{neV}$ the gain in UCN density is not too impressive
(lowest two curves), for $\widetilde{V}=330$\ $\mathrm{neV}$ a magnetic $12$%
-pole with $B_{R}=2.5\ \mathrm{T}$ enhances the saturated low field seeker
UCN density $n_{\infty ,\mathrm{lfs}}$ by a factor $2.3$ from $820$\ $%
\mathrm{cm}^{-3}$ to $1880$\ $\mathrm{cm}^{-3}$. Hence, if the experiment
connected to the source can use the high-energy UCN it provides, the
magnetic multipole is an asset even for a vessel with very good storage
properties. Note that the saturated high field seeker UCN density $n_{\infty
,\mathrm{hfs}}$ for $\widetilde{V}\geq V$ does not depend on $\widetilde{V}$%
, since for them the trapping potential is given by $V-V_{\mathrm{m}R}$ (see
upper part of Fig.\ $2$). Hence, the larger $\widetilde{V}$, the larger will
be the polarization $P_{\infty }$ defined in eq.\ \ref{polarization} (see
also Table $3$). Note also that, as sketched in the lower part of Fig.\ $2$,
for magnetic fields providing a trapping potential stronger than the neutron
optical one ($B_{R}\gtrsim 3.9\ \mathrm{T}$ for the situation shown in Fig.\ 
$4$), $P_{\infty }=1$.

Figure $5$ shows the dependence of saturated UCN density on the multipole
order, for traps with Be-coated cylindrical wall with $f=3\times 10^{-5}$,
respectively $f=2\times 10^{-4}$, each with end windows with a potential of $%
\widetilde{V}=330$\ $\mathrm{neV}$. Again, one notes the positive influence
of higher multipole order on the number of trapped UCN and on the
polarization. Table $3$ quotes values for $n_{\infty ,\mathrm{lfs}}$ and $%
P_{\infty }$ for $n=12$ and for various values of $f$. One can see, for
instance for $f=2\times 10^{-4}$, that the magnetic field enhances $%
n_{\infty ,\mathrm{lfs}}$ by more than a factor five from $230$\ $\mathrm{cm}%
^{-3}$ to $1210$\ $\mathrm{cm}^{-3}$. One also notes that the multipole
field stabilizes the output of the source, by mitigating the influence of
the loss coefficient of the converter wall surface. This includes possible
deterioration of the wall quality with time, which will then also be much
less an issue than without the field.

For application of the source for feeding a magnetic trap, e.g.\ for neutron
lifetime experiments with typical trapping potentials in the range $\left(
50-120\right) $\ $\mathrm{neV}$ (see current projects in refs.\ \cite%
{Ezhov/2014,Salvat/2014,Leung/2014,Ezhov/2009,Picker/2005}), it is
interesting to study the dependence of $n_{\infty ,\mathrm{lfs}}$ on the
upper bound of the trapped UCN energy spectrum. Figure $6$ shows this
dependence, for traps with $n=12$ and again with the cylindrical section
made of Be with $f=3\times 10^{-5}$, respectively $f=2\times 10^{-4}$.
Values are calculated using eq.\ \ref{n_inf} with the potential $\widetilde{V%
}-V_{\mathrm{He}}$ set to different values starting from $60$\ $\mathrm{neV}$
and increased in steps of $20$\ $\mathrm{neV}$. We see that, the lower $%
\widetilde{V}$, the lower will be the magnetic field needed to optimize the
UCN density. The reason is that lowering $\widetilde{V}$ reduces wall
collisional losses due to the energy dependence of $\overline{\mu }\left(
E\right) $ defined in eq.\ \ref{mu} and due to wall hits occuring at a
smaller rate, whereas the effective volume $\gamma ^{\prime }$ decreases
quickly with $V_{\mathrm{m}R}$ for a low-energy UCN spectrum (see Table $1$%
). The solid curves in Fig.\ $6$ tell us that, for the converter vessel
coated with an excellent Be mirror, $f=3\times 10^{-5}$, the multipole
magnet will offer some advantage only for not too low UCN cutoff energy.
However, for a more realistic situation, $f=2\times 10^{-4}$, gains due to
the magnet are rather significant even for low-energy UCN spectra. For
example, for feeding an external trap with trapping depth $60$\ $\mathrm{neV}
$, it will improve $n_{\infty ,\mathrm{lfs}}$ by a factor $2$ at $%
B_{R}\approx 0.8$\ $\mathrm{T}$. With increasing trapping depth the gain
increases, e.g.\ to a factor $3.2$ at $B_{R}\approx 1.8$\ $\mathrm{T}$ for $%
V_{\mathrm{trap}}=120$\ $\mathrm{neV}$.

\begin{figure}[tbp]
\centering\includegraphics[width=1.0\textwidth]{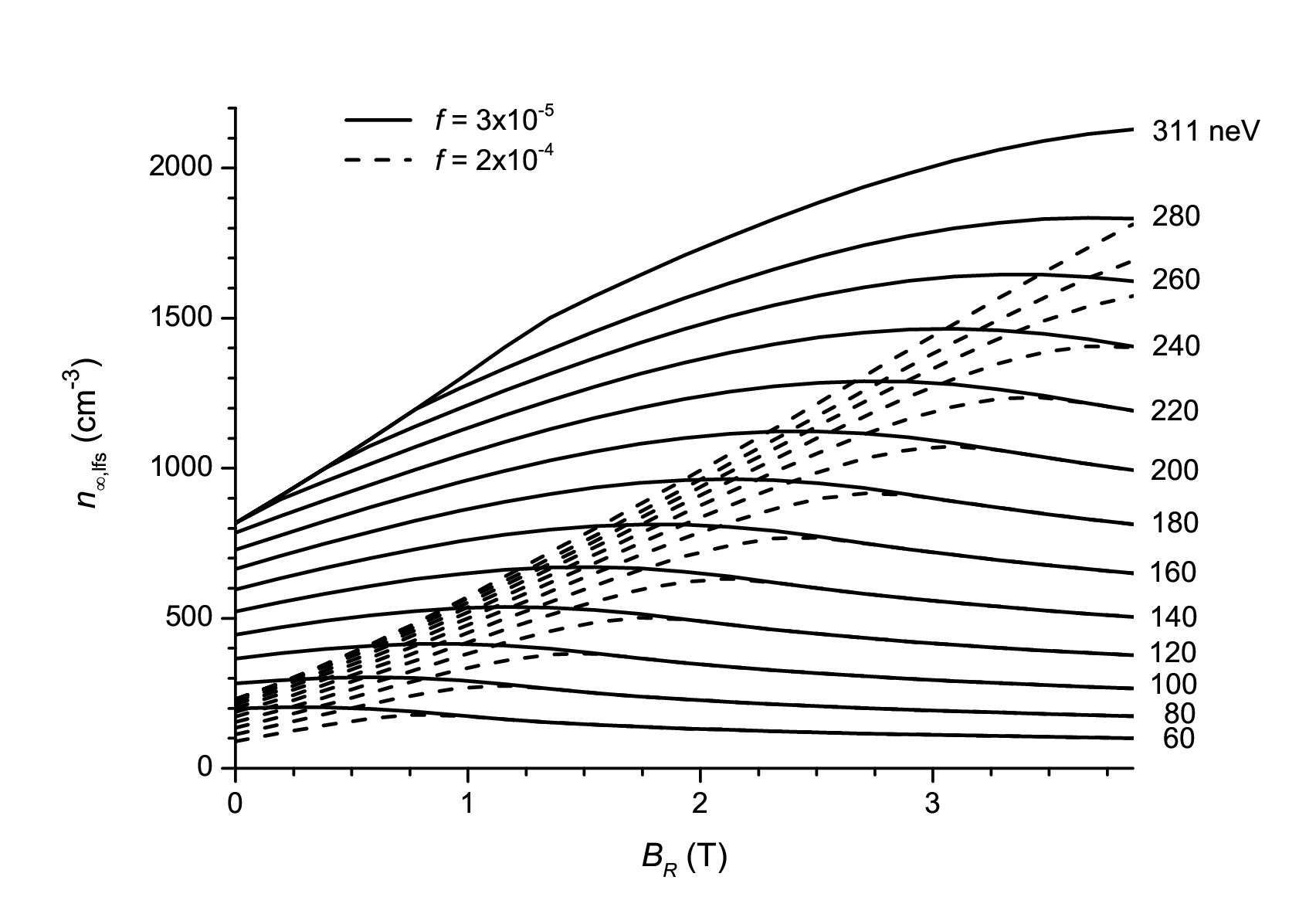}
\caption{Saturated density of low field seeking UCN in converter vessels
with diameter $10$ cm, $V=252$ neV, $T=0.5$ K and surrounded by a $12$-pole
magnet. Solid curves are for $f=3\times 10^{-5}$, dashed ones for $f=2\times
10^{-4}$. $\widetilde{V}-V_{\mathrm{He}}$ is varied between $60$ neV and $311
$ neV. Values are given for $\Phi _{0.89\mathrm{nm}}=10^{9}\ $cm$^{-2}$s$%
^{-1}$nm$^{-1}$ (unpolarized).}
\label{fig:figure6}
\end{figure}
%TCIMACRO{\TeXButton{B}{\begin{table}[tbp] \centering}}%
%BeginExpansion
\begin{table}[tbp] \centering%
%EndExpansion
\begin{tabular}{l|l|l|llll}
$\widetilde{V}$ $\left( \mathrm{neV}\right) $ & $B_{R}$ $\left( \mathrm{T}%
\right) $ & $f$ & $3\times 10^{-5}$ & $1\times 10^{-4}$ & $2\times 10^{-4}$
& $4\times 10^{-4}$ \\ \hline
$330$ & $2.5$ & $n_{\infty ,\mathrm{lfs}}$ $\left( \mathrm{cm}^{-3}\right) $
& $1880$ & $1430$ & $1210$ & $1050$ \\ 
&  & $P_{\infty }$ & $0.89$ & $0.949$ & $0.969$ & $0.981$ \\ \hline
$252$ & $2.5$ & $n_{\infty ,\mathrm{lfs}}$ $\left( \mathrm{cm}^{-3}\right) $
& $1380$ & $1200$ & $1080$ & $980$ \\ 
&  & $P_{\infty }$ & $0.86$ & $0.939$ & $0.965$ & $0.98$ \\ \hline
$\geq 252$ & $0$ & $n_{\infty ,\mathrm{lfs}}$ $\left( \mathrm{cm}%
^{-3}\right) $ & $820$ & $390$ & $230$ & $130$ \\ 
&  & $P_{\infty }$ & $0$ & $0$ & $0$ & $0$%
\end{tabular}%
\caption{Values for the saturated density of low field seeking UCN (see eq.\ \ref{n_inf}) and UCN polarization  
(eq.\ \ref{polarization}), for a converter vessel with $R=5$ cm, $T=0.5$ K, surrounded by a $12$-pole magnet, and 
for several values of $f$. Values are given for an unpolarized differential flux density of 
$\Phi _{0.89\mathrm{nm}}=10^{9}$ cm$^{-2}$s$^{-1}$nm$^{-1}$.}\label{TableKey
copy(2)}%
%TCIMACRO{\TeXButton{E}{\end{table}}}%
%BeginExpansion
\end{table}%
%EndExpansion

\section{Conclusions}

As our analysis shows, a multipole magnet can lead to a large gain in the
saturated density of low field seeking UCN because the presence of the field
reduces the number of neutrons hitting the material walls and reduces the
energy and wall collision rate of those that do. In addition, it acts as a
source-intrinsic UCN polarizer without need to polarize the incident beam
and hence avoiding associated losses. A $12$-pole magnet with field $%
B_{R}=2.5$\ $\mathrm{T}$ on a radius of $R=5$\ $\mathrm{cm}$ seems
technically feasible using standard NbTi superconducting wire technology, as
investigated in an independent study using a finite element code. Based on
results of experimental work done by other groups, a promissing candidate
vessel able to provide a UCN spectrum with exceptionally high cutoff
consists of a Be trap closed off by disks coated with c$^{11}$BN.
Alternative materials are diamondlike carbon with $V$ close to $300$\ $%
\mathrm{neV}$ depending on the abundance of sp$_{3}$ chemical bonds, or
enriched $^{58}$Ni with $\overline{V}=346$\ $\mathrm{neV}$ and a theoretical 
$f=8.6\times 10^{-5}$, which however is magnetic so that UCN depolarization
might be an issue that needs experimental study. We note that, in order to
extract the full benefits, the incoming cold beam will need to be
transported by a supermirror guide, with a top layer deposit of a good UCN
reflecting material with neutron optical potential $V$. An experimental
study of a UCN source prototype involving a converter vessel coated with
such type of mirror is currently underway at the ILL, in preparation of the
UCN source project SuperSUN which will include a $12$-pole magnet around a $%
3 $\ $\mathrm{m}$ long cylindrical converter vessel.

Our benchmark converter is able to provide a saturated low field seeker UCN
density almost as high as an unrealistic, perfect trap with $f=0$ and Be
cutoff, for which one calculates $n_{\infty ,\mathrm{lfs}}=2060$ $\mathrm{cm}%
^{-3}$ when exposed to a neutron beam with differential flux density $\Phi
_{0.89\mathrm{nm}}=10^{9}\mathrm{cm}^{-2}\mathrm{s}^{-1}\mathrm{nm}^{-1}$.
For a pure Be trap equipped with the $12$-pole magnet one calculates $%
n_{\infty ,\mathrm{lfs}}=1380$ $\mathrm{cm}^{-3}$ for $f=3\times 10^{-5}$ as
previously achieved for this material, while for $f$ ten times worse, we
still obtain $n_{\infty ,\mathrm{lfs}}=1110$ $\mathrm{cm}^{-3}$. Without
magnet on the other hand, $n_{\infty ,\mathrm{lfs}}$ would get suppressed by
a factor five, which impressively demonstrates the capability of the
magnetic multipole reflector to mitigate the influence of a poor loss
coefficient $f$ of the converter wall surface. Obviously, also if loss of
quality of the inner converter surface with time will be an issue, the
magnet offers a valuable practical advantage.

Including the high field seekers into the discussion, one first notes that
without magnetic field they are equally well trapped, so that in this case
(and still assuming the usual situation of an unpolarized cold neutron beam
for UCN production) the total UCN density in the source will be a factor two
higher. However, for experiments requiring polarized UCN such as magnetic
traps for precise determination of the neutron lifetime, or the neutron EDM
experiment, this factor is of no use. Values for polarization of the trapped
ensemble of UCN after saturation of the source were quoted in Table $3$ and
are typically well beyond $90\%$ for the system discussed. As obvious from
Fig.\ $2$, low-energy UCN may stay poorly polarized whereas for the
high-energy part of the spectrum the high field seekers, after magnetic
acceleration to the cylindrical wall, will have kinetic energy beyond its
cutoff potential and quickly get lost. Poor polarization is not a problem
for experiments using magnetic traps which can be designed for quick
cleaning out the wrong spin component. For experiments that would profit
from a very high initial polarization one might coat the converter vessel
with a low-loss material with potential $V<V_{\mathrm{m}R}$, which would
lead to nearly $100\%$ UCN polarization since the high field seekers would
stay untrapped. For a system with $V>V_{\mathrm{m}R}$ one may still, if
needed, increase the polarization by delayed extraction of the UCN after
having switched off the saturating neutron beam. The high field seekers will
then quicker leave the trap than the low field seekers due to shorter
trapping time constants. One might also cut out the lowest-energy part of
the spectrum by a vertical UCN guide section with suitably chosen length.

As an important detail not affecting our conclusions we note that, in
addition to the multipole field, it will be necessary to apply a bias field
in the order of some $10$\ $\mathrm{mT}$ along the converter axis to avoid
depolarization in the region around $r=0,$\ where the multipole field is
zero. In addition, we can consider using axial magnetic pinch fields to
increase $V_{\mathrm{trap}}$ and thereby the density of the UCN for an
extended energy spectrum. For extraction, the field at one end needs to be
ramped down, so that an iris type UCN valve might be most appropriate for
this case. An extended UCN spectrum would be interesting if UCN of any
velocity were beneficial, such as in UCN transmission experiments, or in
combination with a phase space transformation by letting the UCN rise
against the gravitational field. Note however that low-loss extraction of
such a UCN spectrum will be a challenge. On the other hand, some studies
might be performed in situ using static pinch fields, such as experiments on
UCN upscattering in superfluid $^{4}$He, for which any increase in UCN
density will be very welcome.


\begin{thebibliography}{99}
\bibitem{Fermi/1936} E. Fermi, Ricerca Scientifica \textbf{7}, 13 (1936).

\bibitem{Luschikov/1969} V.I. Luschikov, Yu.N. Pokotilovsky, A.V. Strelkov,
F.L. Shapiro, Sov. Phys. JETP Lett. \textbf{9}, 23 (1969).

\bibitem{Steyerl/1969} A. Steyerl, Phys. Lett. \textbf{29B}, 33 (1969).

\bibitem{Golub/1991} R. Golub, D.J. Richardson, S.K. Lamoreaux, \textit{%
Ultra-Cold Neutrons} (Adam Hilger, Bristol 1991).

\bibitem{Ignatovich/1990} V.K. Ignatovich, \textit{The Physics of Ultracold
Neutrons} (Oxford Science Publications, Clarendon Press, Oxford, 1990).

\bibitem{Dubbers/2011} D. Dubbers and M.G. Schmidt, Rev. Mod. Phys. \textbf{%
83}, 1111 (2011).

\bibitem{Musolf/2008} M.J. Ramsey-Musolf and S. Su, Phys. Rept. \textbf{456}%
, 1 (2008).

\bibitem{Abele/2008} H. Abele, Prog. Nucl. Phys. \textbf{60}, 1 (2008).

\bibitem{Jenke/2011} T. Jenke, P. Geltenbort, H. Lemmel, H. Abele, Nature
Phys. \textbf{7}, 468 (2011).

\bibitem{Serebrov/2008} A.P. Serebrov, E.B. Aleksandrov, N.A. Dovator et
al., Phys. Lett. B \textbf{663}, 181 (2008).

\bibitem{Ban/2007} G. Ban, K. Bodek, M. Daum et al., Phys. Rev. Lett. 
\textbf{99}, 161603 (2007).

\bibitem{Altarev/2009} I. Altarev, C.A. Baker, G. Ban et al., Phys. Rev.
Lett. \textbf{103}, 081602 (2009).

\bibitem{Jenke/2014} T. Jenke, G. Cronenberg, J. Burgd\"{o}rfer et al.,
Phys. Rev. Lett. \textbf{112}, 151105 (2014).

\bibitem{Serebrov/2010} A.P. Serebrov, O. Zimmer, P. Geltenbort et al., J.
Exp. Theor. Phys. Lett. \textbf{91}, 6 (2010).

\bibitem{Zimmer/2010b} O. Zimmer, Phys. Lett. B \textbf{685}, 38 (2010).

\bibitem{Frank/2008} A.I. Frank, P. Geltenbort, M. Jentschel et al., Phys.
Atomic Nucl. \textbf{71}, 1656 (2008).

\bibitem{Filipp/2009} S. Filipp, J. Klepp, Y. Hasegawa et al., Phys. Rev.
Lett. \textbf{102}, 030404 (2009).

\bibitem{Richardson/1989} D.J. Richardson and S.K. Lamoreaux, Nucl. Instr.
Meth. A \textbf{284}, 192 (1989).

\bibitem{Richardson/1988} D.J. Richardson, A.I. Kilvington, K. Green, S.K.
Lamoreaux, Phys. Rev. Lett. \textbf{61}, 2030 (1988).

\bibitem{Bitter/1987} T. Bitter and D. Dubbers, Phys. Rev. Lett. \textbf{59}%
, 251 (1987).

\bibitem{Arzumanov/2012} S.S. Arzumanov et al., JETP Lett. \textbf{95}, 224
(2012).

\bibitem{Pichlmaier/2010} A. Pichlmaier, V. Varlamov, K. Schreckenbach, P.
Geltenbort, Phys. Lett. B \textbf{693}, 221 (2010).

\bibitem{Paul/2009} S. Paul, Nucl. Instr. Meth. A \textbf{611}, 157 (2009).

\bibitem{Serebrov/2008b} A.P. Serebrov, V.E. Varlamov, A.G. Kharitonov et
al., Phys. Rev. C \textbf{78}, 035505 (2008).

\bibitem{Nico/2005} J.S. Nico, M.S. Dewey, D.M Gilliam, F.E. Wietfeldt et
al., Phys. Rev. C \textbf{71}, 055502 (2005).

\bibitem{Mendenhall/2013} M.P. Mendenhall, R.W. Pattie, Y. Bagdasarova et
al., Phys. Rev. C \textbf{87}, 032501 (2013).

\bibitem{Plaster/2012} B. Plaster, R. Rios, H.O. Back, T.J. Bowles et al.,
Phys. Rev. C \textbf{86}, 055501 (2012).

\bibitem{Mund/2012} D. Mund et al., arXiv:1204.0013 (2012).

\bibitem{Abele/2002} H. Abele, S. Baessler, D. Dubbers et al., Phys. Rev.
Lett. \textbf{88}, 211801 (2002).

\bibitem{Coc/2007} A. Coc, N.J. Nunes, K.A. Olive et al., Phys. Rev. D 
\textbf{76}, 023511 (2007).

\bibitem{Lopez/1999} R.E. Lopez and M.S. Turner, Phys. Rev. D \textbf{59},
103502 (1999).

\bibitem{Mention/2011} G. Mention, M. Fecher, Th. Lasserre et al., Phys.
Rev. D \textbf{83}, 073006 (2011).

\bibitem{Purcell/1950} E.M. Purcell, and N.F. Ramsey, Phys. Rev. \textbf{78}%
, 807 (1950).

\bibitem{Pospelov/2005} M. Pospelov and A. Ritz, Annals Phys. \textbf{318},
119 (2005).

\bibitem{Baker/2006} C.A. Baker et al., Phys. Rev. Lett \textbf{97}, 131801
(2006).

\bibitem{Serebrov/2014} A.P. Serebrov, E.A. Kolomenskiy, A.N. Pirozhkov et
al., Pis'ma v ZhETF \textbf{99}, 7 (2014).

\bibitem{Altarev/2012} I. Altarev et al., Il Nuovo Cimento \textbf{35} C,
122 (2012).

\bibitem{Masuda/2012a} Y. Masuda, K. Asahi, K. Hatanaka et al., Phys. Lett.
A \textbf{376}, 1347 (2012).

\bibitem{Grinten/2009} M.G.D. van der Grinten, Nucl. Instr. Meth. A \textbf{%
611}, 129 (2009).

\bibitem{Altarev/2009b} I. Altarev, G. Ban, G. Bison, K. Bodek et al., Nucl.
Instr. Meth. A \textbf{611}, 133 (2009).

\bibitem{Serebrov/2009b} A.P. Serebrov et al., Nucl. Instr. Meth. A \textbf{%
611}, 263 (2009).

\bibitem{Lamoreaux/2009} S.K. Lamoreaux and R. Golub, J. Phys. G: Nucl.
Part. Phys. \textbf{36}, 104002 (2009).

\bibitem{EDM@SNS/2004} The EDM@SNS neutron EDM experiment,
http://p25ext.lanl.gov/edm/edm.html.

\bibitem{Golub/1994} R. Golub and S.K. Lamoreaux, Physics Reports \textbf{237%
}, 1 (1994).

\bibitem{Karch/2014} J. Karch, Yu. Sobolev, M. Beck et al., Eur. Phys. J. A 
\textbf{50}, 78 (2014).

\bibitem{Lauss/2014} B. Lauss, Phys. Procedia \textbf{51}, 98 (2014).

\bibitem{Piegsa/2014} F.M. Piegsa, M. Fertl, S.N. Ivanov, K.K.H. Leung, M.
Kreuz, P. Schmidt-Wellenburg, T. Soldner, O. Zimmer, Phys. Rev. C \textbf{90}%
, 015501 (2014).

\bibitem{Lauer/2013} T. Lauer, T. Zechlau, Eur. Phys. J. A \textbf{49}, 104
(2013).

\bibitem{Saunders/2013} A. Saunders, M. Makela, Y. Bagdasarova et al., Rev.
Sci. Instrum. \textbf{84}, 013304 (2013).

\bibitem{Masuda/2012} Y. Masuda, K. Hatanaka, S.-C. Jeong et al., Phys. Rev.
Lett. \textbf{108}, 134801 (2012).

\bibitem{Serebrov/2009} A.P. Serebrov, V.A. Mityuklaev, A.A. Zakharov et
al., Nucl. Instr. Meth. A \textbf{611}, 276 (2009).

\bibitem{Korobkina/2007} E.I. Korobkina, B.W. Wehring, A.I. Hawari, A.R.
Young et al., Nucl. Instr. Meth. A \textbf{579}, 530 (2007).

\bibitem{Frei/2007} A. Frei, Y. Sobolev, I. Altarev et al., Eur. Phys. J. A 
\textbf{34}, 119 (2007).

\bibitem{Trinks/2000} U. Trinks, F. J. Hartmann, S. Paul, and W. Schott,
Nucl. Instr. Meth. A \textbf{440}, 666 (2000).

\bibitem{Golub/1975} R. Golub and J.M. Pendlebury, Phys. Lett. \textbf{53A},
133 (1975).

\bibitem{Serebrov/1994} A. Serebrov et al., J. Exp. Theoret. Phys. Lett. 
\textbf{59}, 11 (1994).

\bibitem{Golub/1983-1} R. Golub, K. B\"{o}ni, Z. Phys. B \textbf{51}, 95
(1983).

\bibitem{Altarev/1980} I. Altarev et al., Phys. Lett. A \textbf{80}, 413
(1980).

\bibitem{Kilvington/1987} A.I. Kilvington, R. Golub, W. Mampe, P. Ageron,
Phys. Lett. A \textbf{125}, 416 (1987).

\bibitem{Zimmer/2011} O. Zimmer, F.M. Piegsa, S.N. Ivanov, Phys. Rev. Lett. 
\textbf{107}, 134801 (2011).

\bibitem{Zimmer/2010} O. Zimmer, P. Schmidt-Wellenburg, M. Fertl et al.,
Eur. Phys. J. C \textbf{67}, 589 (2010).

\bibitem{Zimmer/2007} O. Zimmer, K. Baumann, M. Fertl et al., Phys. Rev.
Lett. \textbf{99}, 104801 (2007).

\bibitem{Leung/2014} K. Leung, S. Ivanov, F. Martin et al., Proceedings of
the workshop \textquotedblleft Next Generation Experiments to Measure the
Neutron Lifetime\textquotedblright , Santa Fe, New Mexico, 9 -- 10 November
2012, page 145. World Scientific (2014).

\bibitem{Leung/2009} K.K.H. Leung, O. Zimmer, Nucl. Instr. Meth. A \textbf{%
611}, 181 (2009).

\bibitem{Zimmer/2000} O. Zimmer, J. Phys. G: Nucl. Part. Phys. \textbf{26},
67 (2000).

\bibitem{Huffman/2000} P.R. Huffman, C.R. Brome, J.S. Butterworth et al.,
Nature \textbf{403}, 62 (2000).

\bibitem{Sommers/1955} H.S. Sommers, J.G. Dash, L. Goldstein, Phys. Rev. 
\textbf{97}, 855 (1955).

\bibitem{Dzhosyuk/2005} S.N. Dzhosyuk et al. J. Res. NIST \textbf{110}, 339
(2005).

\bibitem{Golub/1983} R. Golub, C. Jewell, P. Ageron, W. Mampe, B. Heckel, I.
Kilvington, Z. Phys. B \textbf{51}, 187 (1983).

\bibitem{Leung/2013} K.K.H. Leung, Ph.D. thesis, TU Munich, unpublished
(2013).

\bibitem{Ezhov/2005} V.F. Ezhov, A.Z. Andreev, A.A. Glushkov et al., J. Res.
NIST \textbf{110}, 345 (2005).

\bibitem{Serebrov/2000} A. Serebrov et al., Nucl. Instr. Meth A \textbf{440}%
, 717 (2000).

\bibitem{Golub/1979} R. Golub, Phys. Lett. \textbf{72A}, 387 (1979).

\bibitem{Yoshiki/2005} H. Yoshiki, H. Nakai, E. Gutsmiedl, Cryogenics 
\textbf{45}, 399 (2005).

\bibitem{Yoshiki/1994} H. Yoshiki, K. Sakai, T. Kawai, S. Goto'o, Cryogenics 
\textbf{34}, 277 (1994).

\bibitem{McClintock/1978} P. McClintock, Cryogenics \textbf{18}, 201 (1978).

\bibitem{Schmidt-Wellenburg/2009} P. Schmidt-Wellenburg, K.H. Andersen, O.
Zimmer, Nucl. Instr. Meth. A \textbf{611}, 259 (2009).

\bibitem{Baker/2003} C.A. Baker, S.N. Balashov, J. Butterworth et al., Phys.
Lett. A \textbf{308}, 67 (2003).

\bibitem{Korobkina/2002} E. Korobkina, R. Golub, B.W. Wehring, A.R. Young,
Phys. Lett. A \textbf{301}, 462 (2002).

\bibitem{Masuda/2002} Y. Masuda, T. Kitagaki, K. Hatanaka et al., Phys. Rev.
Lett. \textbf{89} (2002) 284801-1.

\bibitem{Ageron/1978} P. Ageron, W. Mampe, R. Golub, J.M. Pendlebury, Phys.
Lett. \textbf{66A}, 469 (1978).

\bibitem{Golub/1977} R. Golub, J. Pendlebury, Phys. Lett. A \textbf{82}, 337
(1977).

\bibitem{Brys/2005} T. Brys, M. Daum, P. Fierlinger et al., Nucl. Instr.
Meth. A \textbf{551}, 429 (2005).

\bibitem{Alfimenkov/1992} V. Alfimenkov et al., JETP Lett. \textbf{55}, 84
(1992).

\bibitem{Atchison/2007} F. Atchison, B. Blau, M. Daum et al., Nucl. Instr.
Meth. B \textbf{260}, 647 (2007).

\bibitem{Atchison/2006} F. Atchison, B. Blau, M. Daum et al., Phys. Rev. C 
\textbf{74}, 055501 (2006).

\bibitem{Atchison/2006b} F. Atchison, B. Blau, M. Daum et al., Phys. Lett. B 
\textbf{642}, 24 (2006).

\bibitem{Grinten/1999} M.G.D. van der Grinten, J.M. Pendlebury, D. Shiers et
al., Nucl. Instr. Meth. A \textbf{423}, 421 (1999).

\bibitem{Sobolev/2010} Yu. Sobolev, Th. Lauer, Yu. Borisov et al., Nucl.
Instr. Meth. A \textbf{614}, 461 (2010).

\bibitem{Ezhov/2014} V.F. Ezhov, A.Z. Andreev, G. Ban et al.,
arXiv:1412.7434 (2014).

\bibitem{Salvat/2014} D.J. Salvat, E.R. Adamek, D. Barlow et al., Phys. Rev.
C \textbf{89}, 052501 (2014).

\bibitem{Ezhov/2009} V.F. Ezhov, A.Z. Andreev, G. Ban et al., Nucl. Instr.
Meth. A \textbf{611}, 167 (2009).

\bibitem{Picker/2005} R. Picker, I. Altarev, J. Br\"{o}cker et al., J. Res.
NIST \textbf{110}, 357 (2005).
\end{thebibliography}
\end{document}